\documentstyle[aps]{revtex}
\begin{document}
\draft 
\title{
Analytical solution of a one-dimensional Ising model with zero
temperature dynamics}
\author{A.\ Prados\footnote{e-mail:prados@cica.es}
and J.\ Javier Brey\footnote{e-mail:brey@cica.es}}
\address{F\'{\i}sica Te\'orica, Facultad de F\'{\i}sica, Universidad
         de Sevilla, Apdo.\ de Correos 1065, E-41080 Sevilla, Spain}
\date{\today}
\maketitle
\begin{abstract}
The one-dimensional Ising model with nearest neighbour
interactions and the zero-temperature dynamics recently considered
by Lefevre and Dean -J.\ Phys.\ A: Math.\ Gen.\ {\bf 34}, L213
(2001)- is investigated. By introducing a particle-hole
description, in which the holes are associated to the domain walls
of the Ising model, an analytical solution is obtained. The result
for the asymptotic energy agrees with that found in the mean field
approximation.
\end{abstract}

\section{Introduction}
\label{sec1}

Ising systems are often used as simple models in many different
fields of statistical physics. In particular, the one-dimensional
Ising model with nearest neighbour interactions and Glauber
dynamics \cite{G63} is ubiquitous.  It shows glass-like behaviour,
including non-exponential relaxation
\cite{An70,Sk83,ByS85,Sp89,ByP93,ByP96a}, a laboratory glass
transition in cooling processes \cite{KyS90,ByP94}, hysteresis
effects in thermal cycling experiments \cite{ByP94}, and aging
effects at low temperatures \cite{Br90,PByS97,LyZ00,GyL00}. When a
weak oscillating external field is applied, the amplitude of the
induced magnetization presents a maximum as a function of the
temperature, a behaviour resembling stochastic resonance
\cite{ByP96b,SyS98}.

Other kind of important Ising systems are those with ``facilitated'' dynamics.
They were introduced in a pioneering work by Fredrickson and Andersen
\cite{FyA84,FyB86}, in the context of the relaxation of structural glasses. In
these models, a  given spin can only flip if its nearest neighbours are in a
certain subset of all their possible configurations. The one-dimensional Ising
model with facilitated dynamics (1SFM) has been extensively studied
\cite{JyE91,EyJ93,RyJ95,FyR96,MyJ99,BPyS99,SyE99}. Although, in general, it is
not possible to find an analytical solution of these facilitated Ising models,
the 1SFM has been exactly solved at zero temperature \cite{FyR96,BPyS99}. Very
recently, the 1SFM has been applied to study the compaction of vibrated granular
systems. A hole (particle) is associated to any spin in its excited (ground)
state. Trying to mimic what is done in real experiments
\cite{KFLJyN95,NKBJyN98}, the tapping process is simulated in the following way.
First, the system freely relaxes to a metastable state ($n=0$) following the
dynamics at zero temperature, in which the processes decreasing the density of
particles are forbidden and any configuration with all the holes being isolated
is an absorbent state of the dynamics \cite{vK92}. Then, the system is tapped,
allowing the density to decrease. Afterwards, a new free relaxation at zero
temperature is done, reaching another metastable state ($n=1$). By repeating
this process, a chain ($n=0,1,2,\ldots$) of metastable configurations is
obtained, with the density of the system increasing as a function of $n$.

An analogous approach to the problem of granular compaction has
been carried out by Lefevre and Dean \cite{LyD01,DyL01}. Their
model is again the one-dimensional Ising system with nearest
neighbour interactions. Nevertheless, the usual Glauber dynamics
\cite{G63} is not a good choice for simulating tapping processes.
This is because there are no metastable states, and at zero
temperature the system always reaches the perfectly ordered
ferromagnetic phase. Then, other zero temperature single spin flip
dynamics was considered, in which only the elementary events
lowering the energy are possible. In this way, all the states
composed of domains of $l$ parallel spins with length $l\geq 2$
are metastable, i.\ e., they are absorbent states for this
dynamics. In such states, there is no spin antiparallel to both of
its nearest neighbours, which are the only spins being able to
flip. With this falling dynamics, the tapping process is simulated
in the same way as described above.

In ref.\ \cite{LyD01}, the authors claimed that the new zero
temperature dynamics does not seem amenable to analytic solution.
In fact, the usual way of solving the Glauber model, by
constructing the hierarchy of equations for the two-spin moments,
does not work. The aim of this paper is to show that it is
possible to solve analytically the time evolution of the model by
going to an equivalent particle-hole description of the Ising
system. The holes are associated to the domain walls separating
arrays of parallel spins (particles). Within this picture, the
metastable states are those with all the holes being isolated, as
in the one-dimensional facilitated Ising model.

The paper is organized as follows. In section \ref{sec2} the Ising
model is introduced, as well as the equivalent particle-hole
description. Section \ref{sec3} is devoted to the analytical
solution of the dynamics. A closed hierarchy of equations is
derived for the probabilities $D_r$ of finding $r+1$ consecutive
holes. The solution of this  hierarchy is obtained by means of a
generating function method. In particular, the asymptotic density
of holes and energy in the metastable state are exactly
calculated. The last section contains some final remarks.

\section{The model}
\label{sec2}

We consider the one-dimensional Ising model with nearest neighbour interactions
and periodic boundary conditions. The hamiltonian of the system is
\begin{equation}\label{2.1}
  {\cal H}=-J \sum_{i=1}^N \sigma_i \sigma_{i+1} \, ,
\end{equation}
where $J>0$ is the coupling constant, having dimensions of energy,
$N$ is the number of spins, and $\sigma_i=\pm 1$. The time
evolution of the system is described by a Markov process with
single spin flip dynamics. Then, the probability
$p(\bbox{\sigma},t)$ of finding the system in configuration
$\bbox{\sigma}\equiv\{\sigma_i\}$ at time $t$ obeys the master
equation
\begin{equation}\label{2.2}
  \frac{d}{dt}p(\bbox{\sigma},t)=\sum_i \left[
  W(\bbox{\sigma}|R_i\bbox{\sigma}) p(R_i\bbox{\sigma},t)-
  W(R_i\bbox{\sigma}|\bbox{\sigma}) p(\bbox{\sigma},t) \right] \,
  .
\end{equation}
Here $R_i\bbox{\sigma}$ is the configuration obtained from $\bbox{\sigma}$ by
flipping the $i$-th spin, and $W(R_i\bbox{\sigma}|\bbox{\sigma})$ is the
transition rate for that process. Following Lefevre and Dean \cite{LyD01}, we
consider a zero temperature dynamics in which only the spin flips lowering the
energy are permitted. Namely, we take
\begin{equation}\label{2.3}
  W(R_i\bbox{\sigma}|\bbox{\sigma})=\frac{\alpha}{4}
  \left( 1-\sigma_{i-1}\sigma_i \right) \left( 1-\sigma_i\sigma_{i+1} \right)
  \, ,
\end{equation}
i.\ e., the transition rate equals $\alpha$ if the $i$-th spin is
antiparallel to both of its nearest neighbours, and vanishes
otherwise. While an exact solution of the usual Glauber dynamics
can be found for all temperatures from the hierarchy of equations
for the two-spin moments \cite{G63}, the same procedure applied to
this ``falling'' dynamics leads to a non-closed set of equations,
since more complex moments, involving three and four spins, appear
in them.

For the above reason, it is convenient to go to an equivalent
particle-hole description of the Ising model. For each site $i$,
we define a new variable
\begin{equation}\label{2.5}
  m_i=\frac{1-\sigma_i\sigma_{i+1}}{2} \, .
\end{equation}
When $m_i=1$ we will say that site $i$ is occupied by a hole,
while if $m_i=0$ we will refer to site $i$ as being  occupied by a
particle. Thus, there is a hole at site $i$ if the spins $i$ and
$i+1$ are antiparallel, while a particle corresponds to spins $i$
and $i+1$ being parallel. It follows that holes are associated to
the domain walls separating arrays of parallel spins (particles).
It is important to note that the number of holes is even for any
configuration with periodic boundary conditions. In terms of the
new variables, the hamiltonian of the system reads
\begin{equation}\label{2.5a}
  {\cal H}=J \sum_{i=1}^N (2 m_i-1)=-JN+2J \sum_{i=1}^N m_i \, .
\end{equation}
We define the average dimensionless energy per spin $\varepsilon$
as
\begin{equation}\label{2.5b}
  \varepsilon (t)\equiv\frac{\langle {\cal H} \rangle_t}{JN}=
  -1+\frac{2}{N}\sum_{i=1}^N \langle m_i \rangle_t \, ,
\end{equation}
where the angular brackets $\langle\ldots\rangle_t$ denote average with
$p(\bbox{\sigma},t)$.

In the particle-hole description, the elementary events involve two adjacent
sites. When the $i$-th spin flips, both $m_{i-1}$ and $m_i$ change their state.
From Eq.\ (\ref{2.3}), the transition rate $\omega(R_{i-1}R_i\bbox{m}|\bbox{m})$
for this process is
\begin{equation}\label{2.6}
  \omega(R_{i-1}R_i\bbox{m}|\bbox{m})=\alpha \, m_{i-1}m_i \, .
\end{equation}
As $\alpha$ only determines the arbitrary time scale, we will set
$\alpha=1$ in the following. The master equation for the
probability $p(\bbox{m},t)$ of finding the system in configuration
$\bbox{m}$ at time $t$ reads
\begin{equation}\label{2.7}
  \frac{d}{dt} p(\bbox{m},t)=\sum_i \left[
  \omega(\bbox{m}|R_{i-1}R_i\bbox{m}) p(R_{i-1}R_i\bbox{m},t)-
  \omega(R_{i-1}R_i\bbox{m}|\bbox{m}) p(\bbox{m},t) \right] \, .
\end{equation}
In the dynamics given by the transition rates in Eq.\ (\ref{2.6}), only two
nearest neighbour holes can turn into two particles. Therefore, it is clear that,
after a long enough time period, the system will become stuck in a
``metastable'' state with all the holes being isolated, i.\ e., surrounded by
two particles. Of course, the reached metastable state will depend on the
initial configuration. This behaviour is reminiscent of the one showed by the
1SFM at $T=0$. In the latter, the system also reaches a metastable state with
all the holes isolated, the specific final state depending on the initial
condition \cite{BPyS99,BPyS00}. Nevertheless, the model considered here and the
1SFM at zero temperature are not equivalent. The elementary dynamical events
occurring in each of them are different. While a particle can be adsorbed on any
empty site with at least one nearest neighbour hole in the 1SFM, in the present
model two particles must be adsorbed simultaneously on two adjacent empty sites
of the one-dimensional lattice.

\section{Analytical solution of the dynamics}
\label{sec3}

In order to analyze the dynamics of the model we focus on the set
of moments
\begin{equation}\label{3.1}
  D_r(t)\equiv\langle m_k m_{k+1} \ldots m_{k+r} \rangle_t \, ,
\end{equation}
with $r\geq 1$. The local density of holes is given by the first term of this
hierarchy,
\begin{equation}\label{3.2}
  D_0(t)=\langle m_k\rangle_t \, ,
\end{equation}
and it is directly related with the energy of the system. From Eq.\
(\ref{2.5b}), it is
\begin{equation}\label{3.2b}
  \varepsilon=-1+2 D_0 \, .
\end{equation}
Note that we are restricting ourselves to spatially homogeneous situations, so
these moments $D_r$ do not depend on their first site $k$. From  its own
definition, the moment $D_r$ gives the probability of finding $r+1$ consecutive
holes starting from a given arbitrary site of the lattice. Using the master
equation (\ref{2.7}) with the transition rates (\ref{2.6}) one gets
\begin{equation}\label{3.3}
  \frac{d}{dt} D_r(t)=-2 D_{r+1}(t) -r D_r(t) \, ,
\end{equation}
for all $r\geq 0$. This hierarchy can be solved by introducing the
generating function
\begin{equation}\label{3.4}
  G(x,t)=\sum_{r=0}^\infty \frac{x^r}{r!} D_r(t) \, ,
\end{equation}
from which all the moments $D_r(t)$ are easily obtained,
\begin{equation}\label{3.4a}
  D_r(t)=\lim_{x\rightarrow 0} \frac{\partial^r}{\partial x^r}
  G(x,t) \, .
\end{equation}
The hierarchy of equations (\ref{3.3}) is equivalent to the
following first order partial differential equation for the
generating function:
\begin{equation}\label{3.5}
  \partial_t G(x,t)=-(2+x) \partial_x G(x,t) \, ,
\end{equation}
which has to be solved with the initial condition
\begin{equation}\label{3.6}
  G_0(x)\equiv G(x,0)=\sum_{r=0}^\infty \frac{x^r}{r!} D_r(0)
  \, .
\end{equation}
By using standard techniques it is easily obtained
\begin{equation}\label{3.7}
  G(x,t)=G_0[(2+x)e^{-t}-2] \, .
\end{equation}
For large times the solution approaches the limit
\begin{equation}\label{3.8}
  G(x,\infty)=G_0(-2) \, ,
\end{equation}
and taking into account Eq.\ (\ref{3.4a}),
\begin{equation}\label{3.9}
  \lim_{t\rightarrow\infty} D_0(t)=G_0(-2) \, , \quad
  \lim_{t\rightarrow\infty} D_r(t)=0 \, , r\geq 1 \, .
\end{equation}
This shows that all the holes are isolated in the metastable state
reached by the system in the long time limit. The probability of
finding $r+1$ consecutive holes, which is equal to $D_r$, vanishes
for $r\geq 1$. On the other hand, the asymptotic value of the
density of holes $D_0(\infty)$ does depend on the initial state,
being always smaller than its initial value. In this sense, it is
worth noting that the hierarchy of equations (\ref{3.3}) has as a
stationary solution any constant value of $D_0$, and $D_r=0$ for
$r\geq 1$.

Now, we are going to particularize the above results for a specific initial
condition. Let us consider that the one-dimensional Ising model is initially at
equilibrium at a certain temperature $T$. Then, the probability that spins $i$
and $i+1$ are antiparallel is given by a function of the temperature $a(T)$,
\begin{equation}
  a(T)=\frac{\exp(-\frac{2J}{k_B T})}{1+\exp(-\frac{2J}{k_B T})}
  \; ,
\end{equation}
where $k_B$ is Boltzmann's constant. The value $a=1/2$ corresponds to
$T\rightarrow\infty$, the system is at $t=0$ in a completely random
configuration. From the form of the hamiltonian in the particle-hole
description, Eq.\ (\ref{2.5a}), it follows that at equilibrium the $m_k$
variables are statistically independent, and
\begin{equation}\label{3.10}
  D_r(0)=a^{r+1} \, .
\end{equation}
By using Eq.\ (\ref{3.6}), one gets the initial condition for the
generating function,
\begin{equation}\label{3.11}
  G_0(x)=a e^{a x} \, ,
\end{equation}
and  Eq.\ (\ref{3.7}) yields
\begin{equation}\label{3.12}
  G(x,t)=a \exp \left\{ a \left[(2+x)e^{-t}-2 \right] \right\} \, .
\end{equation}
Now the expressions for the moments $D_r(t)$ can be directly
calculated from Eq.\ (\ref{3.4a}),
\begin{equation}\label{3.13}
  D_r(t)=a^{1+r} \exp \left[ -rt+ a \left( 2e^{-t}-2\right) \right] \, .
\end{equation}
As discussed in the previous paragraph, all the moments $D_r$ with
$r\geq 1$ vanish in the long time limit. Moreover, the asymptotic
density of holes is
\begin{equation}\label{3.14}
  D_0(\infty)=a e^{-2a} \, ,
\end{equation}
which depends on the initial temperature of the system. Equivalently, from Eq.\
(\ref{3.2b}) the specific dimensionless energy $\varepsilon(\infty)$ in the
metastable state reads
\begin{equation}\label{3.15}
 \varepsilon(\infty)=-1+2a e^{-2a} \, .
\end{equation}
The mean field calculation presented by Lefevre and Dean \cite{LyD01,DyL01}
leads to the same expression (see also \cite{DyL01bis}). The asymptotic energy
$\varepsilon(\infty)$ is maximal for $a=1/2$, i.\ e., when the system starts
from the completely random configuration. This is in contrast with the behaviour
found for the 1SFM. In the latter, if the initial condition is taken as the
equilibrium state for a given value of the temperature $T$, the asymptotic
density of holes is a monotonic function of the initial density of holes
\cite{BPyS99}.

\section{Final remarks}
\label{sec4}

In this work we have studied the time evolution of the
one-dimensional Ising model with the zero temperature dynamics
recently considered by Lefevre and Dean \cite{LyD01,DyL01}. In
this dynamics only transitions decreasing the energy of the system
are allowed. The difference with the usual zero temperature
Glauber dynamics \cite{G63} is that flips of spins parallel to one
of its neighbours and antiparallel to the other one are forbidden.
In other words, the possibility of a diffusive motion of the
domain walls is eliminated.

When trying to compute analytically the evolution of the system,
the usual procedure of constructing the hierarchy of differential
equations for the two spin moments is not useful. We have
introduced a particle-hole description of the Ising model, in
which the holes correspond to the domain walls (or ``defects'')
between arrays of parallel spins. The energy of the system is
directly related to the density of holes. Within this description,
the present model has some similarities with the one-dimensional
facilitated Ising model at zero temperature. Although the possible
elementary dynamical events are different, both systems get
eventually stuck in a metastable state characterized by all the
holes being isolated, i.\ e., surrounded by two particles.

The particle-hole description allows us to solve analytically the
time evolution of the system. A closed hierarchy of equations can
be written for the probability distribution functions $D_r$ of
finding $r+1$ consecutive holes in the system. The general
solution of this hierarchy has been derived. In the long time
limit, the system reaches a state in which $D_r=0$ for all $r\geq
1$, i.\ e., all the holes are isolated. Moreover, the asymptotic
value of the density of holes is not unique, but it depends on the
initial configuration.

Another similarity between the present model and the 1SFM shows up in the
context of granular media. When both models are applied to study tapping
processes, the steady state reached by the system is consistent with Edward's
thermodynamic theory \cite{EyO89,MyE89}. This has been shown numerically in
Lefevre and Dean's model \cite{LyD01}, and analytically for the 1SFM
\cite{BPyS00}. In the latter, an effective dynamics for the tapping process has
been derived. This allows to find explicitly the steady distribution function of
the system, as well as the relationship between the vibration intensity and
Edwards' compactivity. Therefore, it seems worth trying to derive the
corresponding effective dynamics for the tapping process in the model considered
in this paper \cite{PyB01a}.

\section{Acknowledgments}

We acknowledge partial support from the Direcci\'{o}n General de
Investigaci\'{o}n Cient\'{\i}fica y T\'{e}cnica (Spain) through
Grant No. PB98-1124.

%
%

\end{document}